\documentclass[twocolumn,pra,10pt,superscriptaddress,notitlepage,longbibliography]{revtex4-1}
\usepackage{graphicx}
\usepackage{amsmath}
\usepackage{breqn}
\usepackage{natbib}
\usepackage{color}
\usepackage[colorlinks=true,linkcolor=blue,citecolor=blue,urlcolor=blue,pdfborderstyle={/S/U/W 1}]{hyperref}
\usepackage[top= 2.0cm,bottom = 2.0cm,left= 2.0cm,right= 1.8cm]{geometry}

\makeatletter
\let\cat@comma@active\@empty
\makeatother
\begin{document}
\title{\makebox{~}\\[1.0em]\makebox{Minimal scattering entanglement in one-dimensional trapped gases}}
\author{Zachary G. Nicolaou}
\affiliation{Department of Physics and Astronomy, Northwestern University, Evanston, IL 60208, USA}
\author{Bohan Xu}
\affiliation{Department of Physics and Astronomy, Northwestern University, Evanston, IL 60208, USA}
\author{Adilson E. Motter}
\affiliation{Department of Physics and Astronomy, Northwestern University, Evanston, IL 60208, USA}
\affiliation{Northwestern Institute on Complex Systems, Northwestern University, Evanston, IL 60208, USA}

\begin{abstract}
The prospect of controlling entanglement in interacting quantum systems offers a myriad of technological and scientific promises, given the progress in experimental studies in systems such as ultracold trapped gases. 
This control is often challenging because of decoherence, the process by which environmental interactions create spurious entanglements that can destroy the desired entanglement. 
Considering the collisional decoherence that is relevant for quantum measurements utilizing scattering in one-dimensional trapped gases, here we derive a relationship between particle masses and wave packet widths that minimizes the entanglement created during scattering.
We assess the relevance of our results by directly observing  this relationship in the emergent scales of a master equation
for a particle undergoing nonthermal scattering.  
Our relationship is independent of the details of the particle interactions and sheds light on how to design scattering processes that minimize decoherence. \\[1.5em]
\noindent DOI: \href{https://doi.org/10.1103/PhysRevA.99.012316}{10.1103/PhysRevA.99.012316}
\end{abstract}

\maketitle

\fontdimen2\font=2.3pt

From the foundational studies of the EPR paradox to the establishment of Bell's inequalities, quantum entanglement has lain at the heart of theoretical physics for the last century \cite{2001_Bell, 2007_Schlosshauer}.  Entanglement is the essential nonclassical behavior in which different components of a system cannot be described independently even when they are spatially separated.  
One of the greatest challenges in creating and sustaining entanglement is decoherence, the process in which desirable quantum correlations are suppressed by the rapid formation of spurious entanglements between a system and its environment. States that produce minimal entanglement with the environment, so-called pointer states, are of central interest to researchers working to create quantum technologies \cite{2003_Zurek}.

Significant progress has been made in understanding entanglement and decoherence in simplified systems, especially ones with discrete state variables.  {Experimental investigations of entanglement have often focused on spin systems with such discrete variables, but recent progress in ultracold gases has opened a new avenue to explore entanglement in systems with continuous variables \cite{2008_Esteve,Riedel_2012}. In particular, one-dimensional laser-confined ultracold gases have been experimentally realized \cite{Gorlitz_2001}, and the exact solutions of the dynamics of one-dimensional Dirac delta and hard-sphere gases \cite{Lieb_1963,Thacker_1981} offer an appealing setting to observe quantum entanglement that can be easily related to theory. }

The formation of entanglement caused by scattering and the resulting collisional docoherence between a particle and a surrounding environmental gas have been extensively studied. An important early result was that scattering of massless environmental particles off a heavy particle of interest can be described by a master equation in which the off-diagonal elements of the reduced density matrix in the position basis decay with time \cite{1985_Joos_Zeh}. This result has since been refined \cite{1990_Gallis_Feming, 2003_Hornberger_Sipe} and experimentally verified \cite{2003_Hornberger}. However, the decay of the off-diagonal density matrix elements should not continue indefinitely but instead is expected to saturate at scales near the thermal de Broglie wavelength $\lambda_{\mathrm{th}} \equiv h/\sqrt{2\pi mk_BT}$  \cite{1985_Joos_Zeh}.  This saturation was initially incorporated into the master equation formalism by accounting for the recoil due to finite mass in Brownian motion \cite{1983_Caldeira,1995_Diosi}.  Later more general master equations were derived \cite{2000_Vacchini,2006_Adler,2006_Hornberger} and Gaussian solitons were identified as potential pointer states \cite{2009_Busse_Hornberger, 2015_Sorgel_Hornberger}.
\begin{figure*}
\includegraphics[width=2\columnwidth]{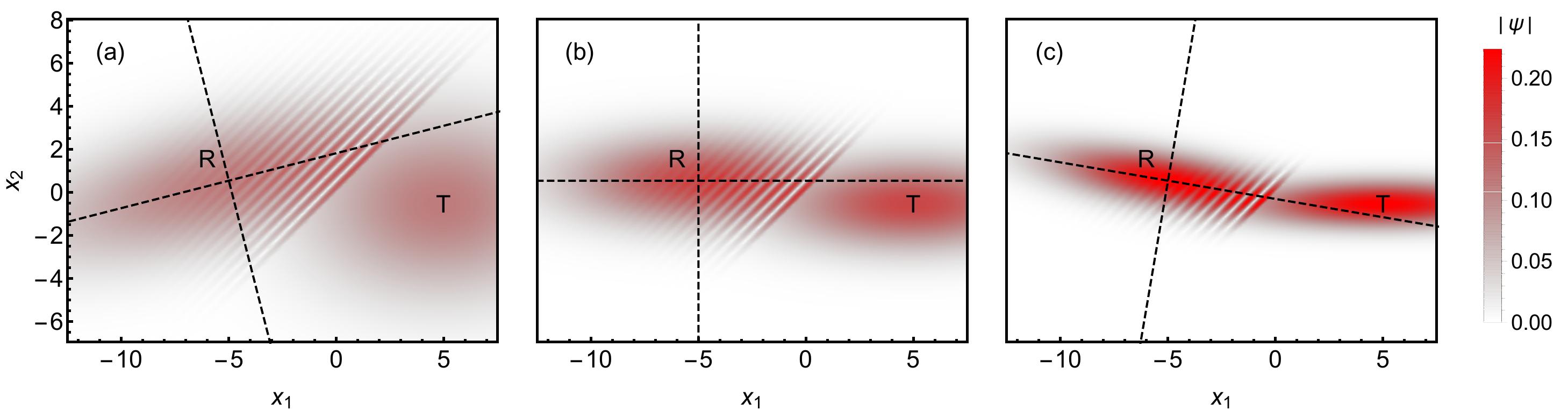}
\caption{Transmitted (T) and reflected (R) wave packets with $m_1=1$ and $m_2=1/9$ and a Dirac delta interaction potential with: (a) a wide mass-width ratio ($\sigma_1=2,~\sigma_2=3$),  (b) the optimal mass-width ratio in Eq.~\eqref{massratio} ($\sigma_1=1,~\sigma_2=3$), and (c) a narrow mass-width ratio ($\sigma_1=1/2,~\sigma_2=3$).   The dashed lines show the orientation of the principle axes of the reflected wave packet. The principle axes are parallel with the coordinate axes when Eq.~\eqref{massratio} is satisfied. \vspace{-1.5em} \label{fig1}}
\end{figure*}

A limitation in this line of research is that the environmental scattering particles have been modeled with the ideal gas density matrix, which is diagonal in the momentum basis.  While this is a reasonable starting point to account for thermal and translationally invariant environments, the problem is that the corresponding states are delocalized,  leading to a density matrix with an intrinsic length scale of $\lambda_{\mathrm{th}}$ in the off-diagonal direction but no intrinsic length scale in the diagonal direction. This density matrix is not relevant to {some nonequilibrium and inhomogneous environments for which we wish to understand entanglement formation.   For example, the dynamical motion of a single impurity particle in an ultracold one-dimensional gas has attracted recent attention. Such an impurity may be a neutral charge \cite{Spethmann_2012} or it may poses spin that differs from the gas \cite{Palzer_2009}. The Josephson effect generating a supercurrent between neighboring one-dimensional traps has also been of theoretical and experimental interest \cite{Didier_2009,Betz_2011}. An impurity particle scattered by an ultracold gas in a trap with geometry that varies laterally or in the presence of a Josephson supercurrent will undergo collisional decoherence because of scattering from the gas, but the ideal gas density matrix is certainly not an appropriate description of this environment.  Indeed, the integrability or near integrability of the dynamics in one-dimensional systems of Dirac delta-interacting bosons precludes rapid thermalization that would lead to an ideal gas environment \cite{Mazets_2010}.} If we wish to consider how collisional decoherence could disrupt engineered entanglement in an experiment studying such scattering, then the density matrix will have to be confined in both the off-diagonal \textit{and} diagonal directions since the trap is itself spatially localized.  Such considerations will be important in the future engineering of quantum technologies. Thus, the question of the impact 
\pagebreak

\noindent of environmental particle wave packet width on the decoherence of a particle in nonthermal environments is pertinent for technological interests. 

Here, we derive a relationship that minimizes undesirable entanglement produced in quantum scattering. 
When restricted to two particles scattering in one-dimension, the relationship takes the form
\begin{equation}
\label{massratio}
\sqrt{r} \sigma_2 =  \sigma_1,
\end{equation}
where $\sigma_1$ and $\sigma_2$ are the widths of the wave packets and $r=m_2/m_1$ is the ratio of the particle masses, $m_1$ and $m_2$. Our mass-width ratio relationship in Eq.~\eqref{massratio} generalizes the natural scale $\lambda_{\mathrm{th}}$ that emerges in a thermal environment to nonthermal environments. In addition to illustrating and deriving our main result, 
we also derive a master equation to show how Eq.~\eqref{massratio} constrains the evolution of the density matrix in a nonthermal environment. For notational convenience, all quantities in the remainder of the text are presumed to be nondimensionalized by appropriate scales unless otherwise noted.  In particular, masses are expressed in terms of a reference $m_{\mathrm{amu}}$ of one atomic mass unit, the thermal de Broglie wavelength  $\lambda_{\mathrm{th}}$ of the mass $m_{\mathrm{amu}}$ is used to nondimensionalize length scales, the thermal momentum $m_{\mathrm{amu}}v_{\mathrm{th}}= \sqrt{m_{\mathrm{amu}}k_B T}$ is used to nondimensionalize momenta, and $t_c=\lambda_{\mathrm{th}}^2m_{\mathrm{amu}}/\hbar$ is used to nondimensionalize time.    

We first consider a single scattering event and suppose the wavefunction is initially an incident wave packet $\psi_{\mathrm{inc}}(x_1,x_2,t)$.  The product form $\psi_{\mathrm{inc}}(x_1,x_2,t)=\psi_{\mathrm{inc}}^1(x_1,t)\psi_{\mathrm{inc}}^2(x_2,t)$ implies that the particles are initially unentangled. In the scattering process, the incident wave packet evolves into transmitted and reflected wave packets,
\begin{equation}
\label{scatter}
\psi_{\mathrm{inc}}(x_1,x_2,t) \xrightarrow[t\to \infty]{} \psi_{\mathrm{refl}}(x_1,x_2,t) +  \psi_{\mathrm{trans}}(x_1,x_2,t).
\end{equation} 
Entanglement is created in this process first because the centers of the reflected and transmitted wave packets differ. If one detects that the first particle reflected off (or transmitted through) the second, then the position and momentum of the second particle are constrained. This form of entanglement resulting from recoil is clearly unavoidable, and is reflected by the fact that the sum in Eq.~\eqref{scatter} does not decompose into a product of independent functions of $x_1$ and $x_2$ even when the incident wave packet does. We will see that the transmitted wave packet itself automatically maintains a product form $\psi_{\mathrm{trans}}(x_1,x_2,t)=\psi_{\mathrm{trans}}^1(x_1,t)\psi_{\mathrm{trans}}^2(x_2,t)$ when the incident wave packet does. On the other hand, the reflected wave packet cannot generally be written in a product form, which represents an additional form of entanglement produced by the scattering. However, under the usual assumption that the incident wave packet is unentangled, the product form $\psi_{\mathrm{refl}}(x_1,x_2,t)=\psi_{\mathrm{refl}}^1(x_1,t)\psi_{\mathrm{refl}}^2(x_2,t)$ does follow when Eq.~\eqref{massratio} is satisfied. Thus, if Eq.~\eqref{massratio} holds, then the spurious entanglement produced in the reflected wave packet is eliminated and the entanglement produced by the scattering is minimized.

\begin{figure}[h]
\includegraphics[width=\columnwidth]{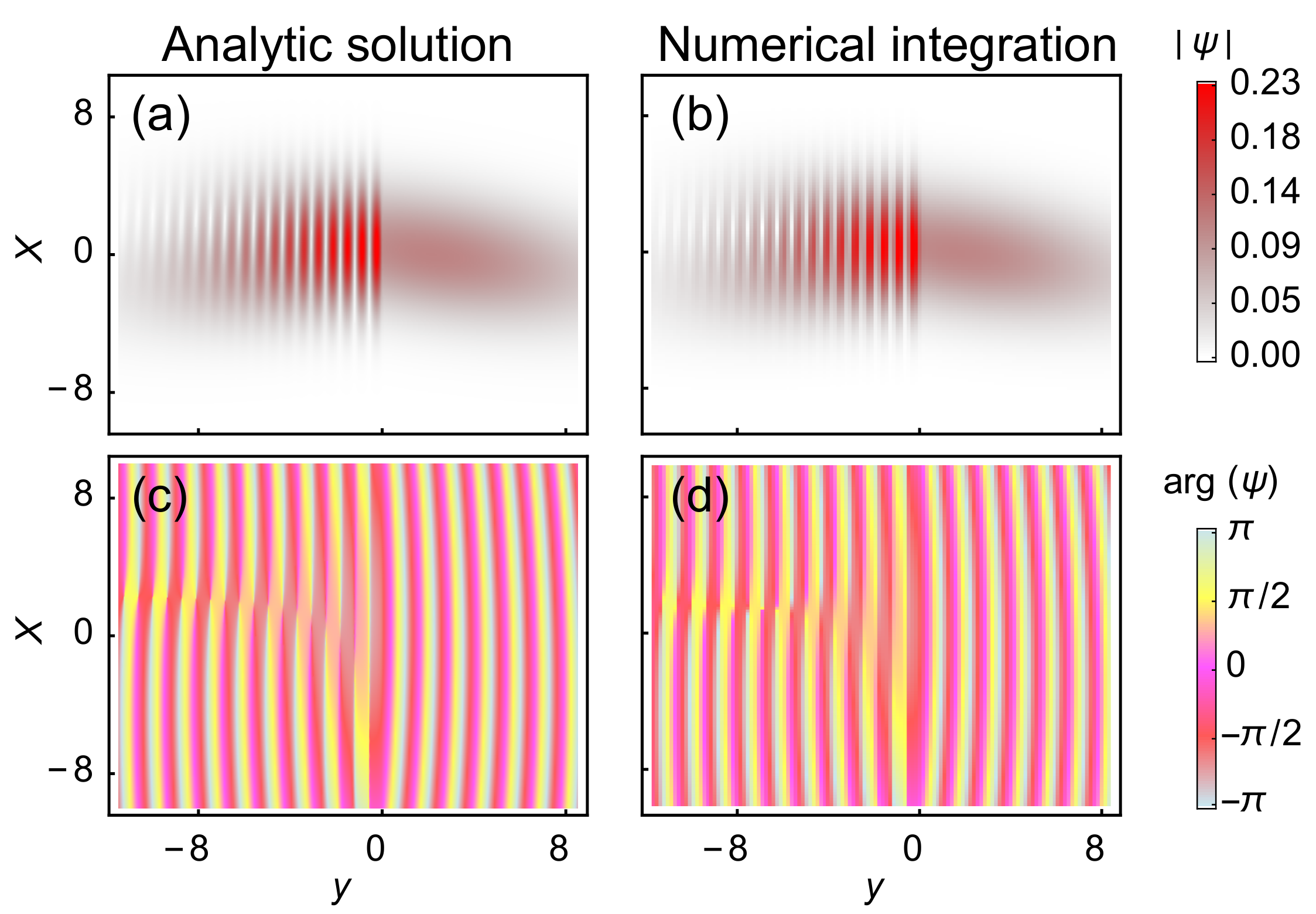}
\caption{[(a) and (b)] Wavefunction amplitude and  [(c) and (d)] wavefunction phase for the analytic solution [(a) and (c)] and the direct numerical integration [(b) and (d)] of the Dirac delta scattering in center-of-mass coordinates, $X=\left(m_1x_1+m_2x_2\right)/\left(m_1+m_2\right)$ and $y=x_2-x_1$. The parameters are the same as in Fig.~\ref{fig1}(a).   \label{figs1}}
\end{figure}

For concreteness, consider the Hamiltonian $H = p_1^2/2m_1 + p_2^2/2m_2 + a \delta(x_2-x_1)$, where $\delta$ is the Dirac delta function.  The evolution of the wave function $\vert \psi \rangle \equiv \int {\mathrm d}x_1 {\mathrm d}x_2 ~ \psi(x_1,x_2,t) \vert x_1 \rangle \otimes \vert x_1 \rangle$ is governed by the Schr\"odiner equation, $i {\mathrm d}\left\lvert \psi \right \rangle/{\mathrm d}t = H \left\lvert \psi \right \rangle$. Utilizing piecewise plane waves that diagonalize the Hamiltonian and the known Hilbert transform of the Gaussian, we derive an exact solution involving the error function that asymptotically approaches Gaussian wave packets in the $t\to \pm\infty$ limits. Figure \ref{fig1}(a) shows the solution for three different choices of wave packet widths during the scattering---the explicit analytic form of this solution is not essential here and is presented in the Appendix \ref{desc}. This analytic solution has been verified against direct numerical integration of the Schr\"{o}dinger equation. As shown in Fig.~\ref{figs1}, the analytical solution and the numerical integration are in excellent agreement. 

The principle axes shown are the coordinates that diagonalize the quadratic exponential in the wavefunction.  These axes can be defined asymptotically for arbitrary interaction potentials in the large time limit, but can also be defined during scattering given the exact solution for this interaction potential. The key feature of the wave packets that satisfy Eq.~\eqref{massratio} is that the incident wave packet is separable in both the laboratory and center-of-mass coordinates. 
The asymptotic angle $\theta$ between the principle axes of the reflected wave packet and the coordinate axes can be analytically derived for the Dirac delta scattering and is shown in Fig.~\ref{fig2}(a).  When Eq.~\eqref{massratio} is satisfied, the angle is zero and the only entanglement produced by the scattering is the recoil form.

To illustrate that the alignment of the principle axes of the reflected wave packet does actually minimize the entanglement created by the scattering,  we consider the entanglement entropy $S = \mathrm{Tr } \left(\rho_1 \ln \rho_1\right)$, where $\rho_1 = \mathrm{Tr_2} \lvert \psi \rangle \langle \psi \rvert $ is the reduced density matrix obtained by taking the partial trace over the second particle. This measure quantifies the degree to which the reduced density matrix fails to be a pure state.  The entanglement entropy was computed along the trajectories of the three wave packets shown in Fig.~\ref{fig1}, and the evolution is shown in Fig.~\ref{fig2}(b).  The entropy rises from zero during the scattering before approaching an asymptotic value $S_{\infty}$ as $t\to\infty$ after scattering. Figure \ref{fig2}(c) shows the asymptotic entropy $S_{\infty}$ as a function of the ratio of the wave packet widths for a family of scattering events.  It is clear that the entropy is minimized exactly when the principle axes of the reflected wave packet are the laboratory coordinate axes (i.e., when $\theta=0$).  It is remarkable that the asymptotic entanglement entropy is not only minimized when Eq.~\eqref{massratio} is satisfied, but also that the value of that minima appear to conincide for different values of $r$. 

{We next consider possible generalizations beyond two-particle, one-dimensional scattering.  Consider $N$ particles with mass $m_i$, position $\mathbf{x}_i$, and momentum $\mathbf{p}_i$ evolving under a Hamiltonian 
$H = \sum_i \frac{\mathbf{p}_i^2}{2m_i} + V\left(\left\{\mathbf{x}_i - \mathbf{x}_1\right\}\right)$. Here, the interaction potential $V$ depends only on the relative separation between particles, and thus the system is invariant under translations, so that the total momentum is conserved. Because of the translational invariance of the Hamiltonian, the center-of-mass coordinates are physically significant. The coordinate transformation to the center of mass is given by $\mathbf{X} \equiv \sum_j m_j \mathbf{x}_j / M, \quad \mathbf{y}_j \equiv \mathbf{x}_j - \mathbf{x}_1$, 
and the inverse transformation is $\mathbf{x}_j = \mathbf{y}_j + \mathbf{X} - {\sum_k m_k \mathbf{y}_k}/{M},$
where $M=\sum_j m_j$ and we take $\mathbf{y}_1 \equiv \mathbf{0}$ for notational simplicity.  

In the center-of-mass coordinates, the Schr\"odinger equation is
\begin{align}
\label{schrondinger2}
i \frac{\partial \psi}{\partial t} &= -\frac{1}{2M} \nabla_X^2 \psi - \frac{1}{2m_1}\sum_{j,k} \nabla_{y_j}\cdot \nabla_{y_k} \psi \nonumber \\
&\quad -\sum_j \frac{1}{2m_j}\nabla_{y_j}^2 \psi  + V(\{\mathbf{y}_j\})\psi. 
\end{align}
Since Eq.~\eqref{schrondinger2} is autonomous with respect to $\mathbf{X}$ and the derivative $\nabla_X$ appears alone in the first term only, it is possible to seek separated solutions of the form  $\psi\left(\mathbf{X}, \{\mathbf{y}_j\},t\right) = \phi_X(\mathbf{X},t)\phi_y(\{\mathbf{y}_j\},t)$.
Substituting this separated solution 
into Eq.~\eqref{schrondinger2} and dividing by $\phi_X\phi_y$, it is possible to perform a separation of variables.  Thus $\phi_X$ and $\phi_y$ satisfy independent Schr\"odinger equations.

The separated solutions maintain their product form for all time, and thus the only requirement to ensure a product form in the center-of-mass coordinates in Eq.~\eqref{schrondinger2} is that the initial condition be in this product form.  Suppose the initial condition is Gaussian in the $(\mathbf{x}_1,\cdots,\mathbf{x}_N)$ laboratory coordinates,
$\psi_0(\left\{\mathbf{x}_j\right\}) = A \exp\left( \sum_{k,\ell, m}\alpha_{\ell m}^k x_k^{\ell} x_k^{m} + \cdots \right)$, where $A$ is a normalization constant, $\alpha^k$ is a negative-definite, symmetric matrix encoding the spread of the $k$th particle and the indices $\ell$ and $m$ denote Cartesian coordinates.  
Here, the additional   \onecolumngrid

\begin{figure}[hb]
\includegraphics[width=0.85\columnwidth]{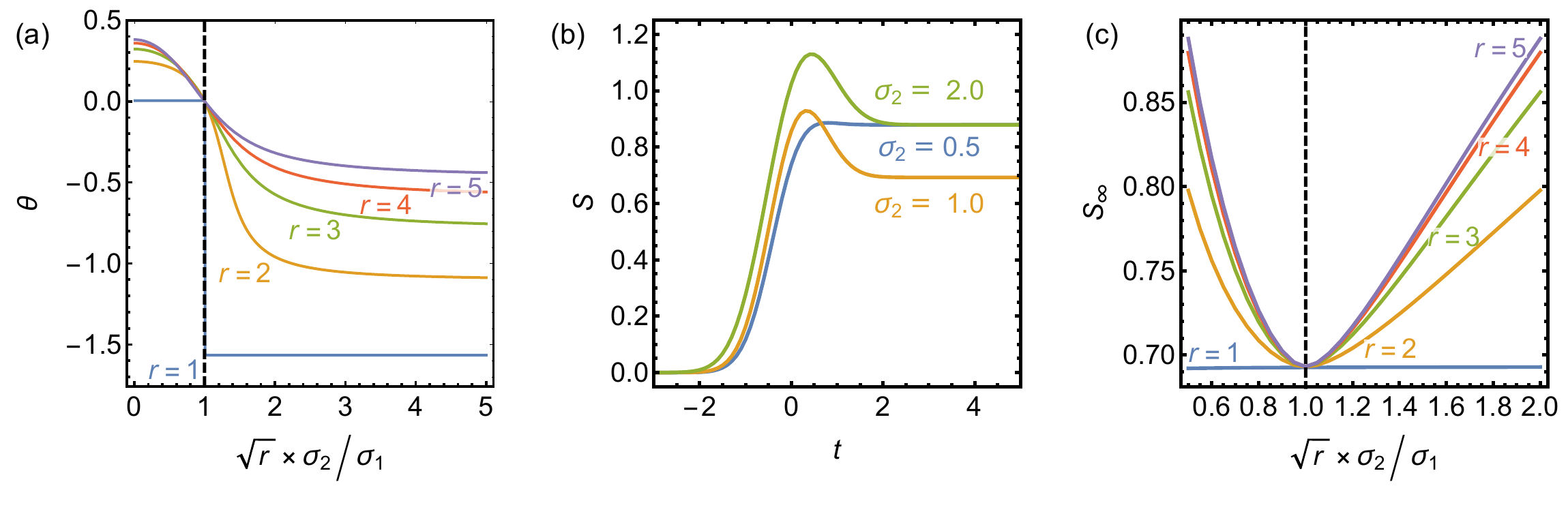}
\vspace{-2em}
\caption{ Entanglement measures for scattering events. (a)  Asymptotic angle $\theta$ of the principle axes of the reflected wave vs.\ width ratio for various choices of the mass ratio, where $\theta$ is zero when Eq.~\eqref{massratio} is satisfied (dashed line). (b) Entanglement entropy $S$ vs. time for the scattering states in Fig.~\ref{fig1}; $S$ increases during scattering and approaches an asymptotic values after scattering. (c) Asymptotic entanglement entropy $S_{\infty}$ vs.\ width ratio for the mass ratio as in Fig.~\ref{fig1}; the entanglement produced in the scattering is minimized exactly when the principle axes of the reflected wave packet coincide with the coordinate axes.  \label{fig2}}
\end{figure}

\twocolumngrid
\clearpage
\pagebreak 

\noindent  terms denoted by $\cdots$ are linear in the $x_k^\ell$ coordinates and encode the initial positions and momenta of the particles.  In the center-of-mass coordinates, the exponential will generally contain cross terms like $X^\ell y_j^m$ that are not in the required product form for the separated solutions.  Employing the inverse transformation, it follows that  the coefficient of the nonproduct term $X^\ell y_j^m$ is $2\left(\alpha_{\ell m}^j  - m_j\sum_k \alpha_{\ell m}^k/M \right)$. If all these coefficients vanish or, equivalently, if 
\begin{equation}
{m_j}/{\alpha^j_{\ell m}} = {m_k}/{\alpha^k_{\ell m}} ~ \forall ~ 1\leq j,k \leq N~\&~ 1\leq \ell, m \leq D, \label{massratio2}
\end{equation}
then all the nonproduct terms will vanish and the initial state will be in a product form in both the center-of-mass coordinates and the laboratory coordinates.  Thus, Eq.~\eqref{massratio2} generalizes Eq.~\eqref{massratio} by the requirement that the solution be separable in both the center-of-mass coordinates and the laboratory coordinates.
\begin{figure*}
\includegraphics[width=1.95\columnwidth]{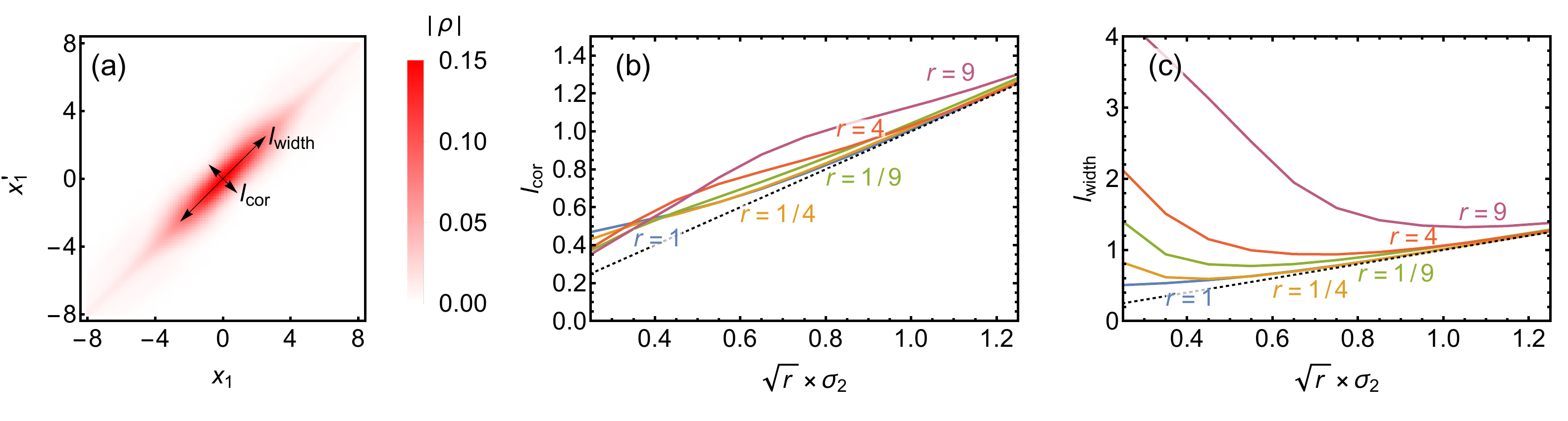} 
\vspace{-1.2em}
\caption{(a) Steady-state density matrix of an initially localized state with $\sigma_2=0.125$ and $r=4$.  The ensemble width $l_{\mathrm{width}}$ describes the classical spread in the possible positions of the particle, while the correlation length $l_{\mathrm{cor}}$ describes the scale of the quantum correlations. 
These length scales are quantified as the full width at half maximum. (b) Steady state $l_{\mathrm{cor}}$ and (c) $l_{\mathrm{width}}$ vs.\ $\sqrt{r}\sigma_2$ for a variety of mass ratios $r$, where the identity function is shown as a reference.  The correlation length is always comparable to $\sqrt{r}\sigma_2$, while the ensemble width approaches $\sqrt{r}\sigma_2$ for large $\sigma_2$.  \vspace{-1em} \label{fig3}}
\end{figure*}}

The asymptotic behavior of an initial localized wave packet can be described in the ${\cal S}$ matrix formalism \cite{1995_Weinberg}.  Suppose that the initial state is exponentially localized in momentum space (and thus position space as well), 
\begin{equation}
\lvert \psi \rangle = \int {\mathrm d}^Dp^1\cdots{\mathrm d}^Dp^N~ \lvert \{\mathbf{p}^j\}_{\mathrm in} \rangle {\mathrm e}^{\sum_{k,\ell, m}\alpha_{\ell m}^k p_k^{\ell} p_k^{m} + \cdots}, 
\end{equation}
where $\lvert \{\mathbf{p}^j\}_{\mathrm in} \rangle$ are the scattering in states that asymptotically approach a free particles with momenta $\mathbf{p}^j$ as $t\to -\infty$. The $t\to+\infty$ asymptotic behavior of $\langle \{\mathbf{x}^j\} \rvert \psi \rangle$ can be found by expressing $\langle \{\mathbf{x}\}^j \rvert$ in terms of scattering out momentum states $\langle \{\mathbf{p}^j\}_{\mathrm{out}} \rvert$ using the fact that they are asymptotically free particles and noting that $\langle \{\mathbf{p'}^k\}_{\mathrm{out}} \rvert \{\mathbf{p}^j\}_{\mathrm{in}} \rangle = {\cal S}(\{\mathbf{p}^j\},\{\mathbf{p'}^{k}\})$ where $\cal S$ is the $\cal S$ matrix. It follows that the wavefunction in the spatial coordinate basis as $t\to \infty$ can be written
\begin{align}
\psi(\mathbf{x}_j,t) &= \int {\mathrm{d}}{\mathbf{p}}~{\cal S}(\{\mathbf{p}^j\},\{\mathbf{p'}^{k}\})\exp \Big[i\sum_j\mathbf{{p'}^j}\cdot\mathbf{x}^j \nonumber \\
&\quad- it\sum_j{\mathbf{p}^j}^2/2m_j +\sum_{k,\ell, m}\alpha_{\ell m}^k p_k^{\ell} p_k^{m} + \cdots \Big] , 
\end{align}
where $ {\mathrm{d}}{\mathbf{p}}= {\mathrm d}^Dp^1 {\mathrm d}^D{p'}^1\cdots{\mathrm d}^Dp^N{\mathrm d}^D{p'}^N $ is shorthand for integration over all momenta variables. The $\cal S$ matrix will generally contain energy conserving Dirac delta factors that will result in reflected and transmitted wave packets in this integration.  Furthermore, the asymptotic behavior of the remaining exponential integral can be found with the method of steepest descent (assuming the $\cal S$ matrix does not have other exponential dependence on its arguments or poles corresponding to bound states), and will result in exponential localization of the wavefunction in position space. Even in one dimension, the asymptotic form will not generally be strictly Gaussian when the initial state is because of the $\cal S$ matrix contribution, although it will be exponentially localized. In more than one dimension, the corresponding localization will be to a spherical shell in the reflected wave packet, and a more explicit partial wave analysis is called for.  However, as we argued previously, this wavefunction will necessarily be separable in the center-of-mass coordinates when Eq.~\eqref{massratio2} is satisfied. For the one-dimensional scattering case above, we saw a clear connection between separability in the center of mass coordinate system and minimization of entanglement produced by scattering. It remains an open question to what extent this center-of-mass separability continues to minimize entanglement in this more general setting, but we conjecture that the center-of-mass seperable states are indeed also the entanglement entropy minimizing states in higher dimensions as well.  

To assess the relevance of Eq.~\eqref{massratio} to collisional decoherence in nonthermal environments 
{relevant for trapped one-dimensional ultracold gases}, we next derive a master equation for a one-dimensional particle subject to repeated scattering by environmental particles of fixed width.  Following the derivation in Ref.~\cite{2006_Hornberger}, we consider scattering that occurs at a rate given by the quantum collision rate operator $\Gamma = \gamma\int {\mathrm d}p_1 {\mathrm d}p_2 ~ |p_1^{\mathrm{rel}}(p_2)| \sigma_{\mathrm{tot}}(p_1^{\mathrm{rel}}(p_2)) |p_1,p_2\rangle\langle p_1,p_2|$
with $\sigma_{\mathrm{tot}}$ denoting the total cross section, $\gamma$ a constant related to the number of environmental particles, and relative momentum given by $p_1^{\mathrm{rel}}(p_2) \equiv \left({m_2p_1} -{m_1p_2}\right)/\left({m_1+m_2}\right)$. We denote the $\cal S$ matrix elements describing the scattering between the system particle and environmental particles with initial and final total momentum $P_{i,f}$ and relative momenta $p_{i,f}$, respectively, by
\begin{equation}
{\cal S}_{i f} = \delta(P_i-P_f)\left[\delta(p_i-p_f) - \frac{if(p_i,p_f)}{\pi}\delta\left(p_i^2-p_f^2\right)\right],
\end{equation}
 where $f(p_i,p_f)$ is the scattering amplitude.

The master equation is given by ${{\mathrm d} \rho}/{{\mathrm d} t} = {\mathrm{Tr}}_{\mathrm{env}}\left(\Gamma^{1/2} {\cal S} \left(\rho\otimes\rho_{\mathrm{env}} \right) {\cal S}^\dag \Gamma^{1/2} \right)$, where $\rho_{\mathrm{env}}$ is the density matrix of the environmental particles.
Here, unlike in Ref.~\cite{2006_Hornberger}, evaluating the trace over the environment in the momentum basis is possible without regularization:
\begin{widetext}
\begin{align}
\label{master}
& \frac{\partial \rho(p_1,p_1')}{\partial t} = \int{\mathrm d}p_2~ K_{-}(p_2,p_1,p_1')\rho\left[\frac{2p_2+(1-r)p_1}{1+r}, \frac{2p_2+(1-r)p_1'}{1+r}\right]\rho_{\mathrm{env}}\left[\frac{2rp_1-(1-r)p_2}{1+r}, \frac{2rp_1'-(1-r)p_2}{1+r}\right] \nonumber \\
&\hspace{6em} -\int {\mathrm d}p_2~ K_{+}(p_2,p_1,p_1')\rho(p_1,p_1')\rho_{\mathrm{env}}(p_2,p_2) - \frac{i\left(p_1^2-{p_1'}^2\right)}{2 m_1}\rho(p_1,p_1') ,  \\
& K_{\pm}(p_2,p_1,p_1') \equiv (2\pi)^2\gamma f\left(p_1^{\mathrm{rel}}(p_2),\pm p_1^{\mathrm{rel}}(p_2)\right)f\left({p_1'}^{\mathrm{rel}}(p_2),\pm{p_1'}^{\mathrm{rel}}(p_2)\right)\sqrt{\frac{\sigma_{\mathrm{tot}}\left[p_1^{\mathrm{rel}}(p_2)\right]\sigma_{\mathrm{tot}}\left[{p_1'}^{\mathrm{rel}}(p_2)\right]}{|p_1^{\mathrm{rel}}(p_2)||{p_1'}^{\mathrm{rel}}(p_2)|}}. \label{master2}
\end{align}
\end{widetext}

To explore the effect of scatterer length scale for environments other than the ideal gas, we consider an artificial environment consisting of Gaussian particles near the origin, with momentum space density matrix
\begin{equation}
\label{rhoenv}
\rho_{\mathrm{\mathrm{env}}}(p_2,p_2') = \sqrt{\frac{2\sigma_2^2}{\pi}}\mathrm{e}^{-\sigma_2^2\left(p_2^2+{p_2'}^2\right)}.
\end{equation}
{Such an environmental density matrix could be applicable in describing an impurity particle under the influence of scattering by a trapped one-dimensional ultracold gas.  For example, the trap geometry can vary on a lateral length scale comparable to the thermal de Broglie wavelength of the gas. First, this is because the temperature is small, and thus the de Broglie wavelengths of all the particles are large. Secondly, the current trap designs permit micro- or nano-scale variations in the trap potential, so that the environmental conditions experienced by an impurity atom can vary over short length scales.  In this case, the translationally invariant ideal gas density matrix may be less appropriate than the density matrix in Eq.~\eqref{rhoenv}, which explicity breaks translational invariance. Another possible application of  Eq.~\eqref{rhoenv} could be to an impurity particle in the vicinity of a Josephson supercurrent in a dilute one-dimensional ultracold gas, since we expect a constant source of localized and dilute gas particles to be present near the tunneling point.  In these examples, note that the mass- and length-scales of the gas particles and the impurity particle can be comparable, so that common approximations of heavy system particles and light environmental particles employed in previous work focused on collisional decoherence may not be applicable.}

Figure \ref{fig3} shows results from the numerical integration of Eqs.~\eqref{master} and \eqref{master2} using the environment in Eq.~\eqref{rhoenv} with Dirac delta interaction potentials 
for a variety of mass ratios $r$ and environmental particle widths $\sigma_2$.  For the Dirac delta potential, the $\cal S$ matrix (and the resulting scattering amplitude $f$ and total cross section $\sigma_{\mathrm{tot}}$) can be determined through scattering theory or, equivalently, by summing the 
\pagebreak 

\noindent Dyson series \cite{1995_Weinberg},
\begin{equation}
\label{deltaS}
{\cal S}_{i f} = \delta(P_i-P_f)\left[\delta(p_i-p_f)  + i\frac{a\mu|p_i|\delta\big(\frac{p_i^2-p_f^2}{2}\big)}{|p_i|-ia\mu}\right],
\end{equation}
so that $f(p_i,p_f) = \frac{-2\pi a\mu|p_i|}{|p_i|-ia\mu}$ and $\sigma_{\mathrm{tot}}(p_i) = \frac{2a^2\mu^2}{|p_i|^2+a^2\mu^2}$.   We fix the intrinsic momentum scale $a\mu = 1$, and, to ensure comparable scattering rates, the particle density constant $\gamma$ scales with $\sigma_2$ as $\gamma=10\times\sqrt{\sigma_2}$.  Then, starting from a Gaussian initial condition with width $\sigma_1 = \sqrt{r}\sigma_2$, the density matrix is found to converge to a steady state after approximately $t>10$, as shown in the position basis in Fig.~\ref{fig3}(a). To demonstrate the importance of the mass-width ratio relationship in Eq.~\eqref{massratio}, we show how the length scales $l_{\mathrm{width}}$ and $l_{\mathrm{cor}}$ vary with $r$ and $\sigma_2$ in Figs.~\ref{fig3}(b) and \ref{fig3}(c).  We find that the correlation length is close to the predicted width $\sqrt{r}\sigma_2$ in all cases, while the ensemble width approaches this length as $\sigma_2$ increases.

In summary, we have studied the entanglement created by scattering particles in a nonthermal environments.  
We recognized two forms of entanglement created by scattering---the first based on recoil and the second based on the deformation of the principle axes of the reflected wave packet---and found that the second form can be eliminated when our mass-width ratio relationship in Eq.~\eqref{massratio} is satisfied.  To assess the relevance of this relationship to collisional decoherence, we derived a master equation for a particular environment of scatterers and found that the emergent scales of its steady states followed the mass-width ratio relation. 
While we focused on one-dimensional scattering between pairs of particles, we noted possible generalization to more dimensions and many-body scattering, and we suggested that separability in multiple physically important coordinate systems, like the center-of-mass and laboratory coordinate systems here, may be a more general feature of entanglement minimizing states. On the other hand, generalization to many particles in the one-dimensional delta-interacting gas may be possible with the exact Bethe ansatz \cite{Lieb_1963} or inverse scattering solutions \cite{Thacker_1981}.  Given the contrast between the complexity of the Bethe ansatz and the simplicity of the two particle scattering here, however, we leave this possibility open to future research. 

\acknowledgments{The authors acknowledge helpful discussions with Pallaboratory Goswami. This work was supported by Northwestern University's WCAS SRG Program and the Simons Foundation (Award No. 342906).
}

{
\appendix
\renewcommand\appendixname{APPENDIX}
\renewcommand{\thesection}{\hspace{-0.2em}}
\renewcommand{\theequation}{A\arabic{equation}}
\section{ANALYTIC SOLUTION DESCRIPTION}
\label{desc}
The analytic solution to the Dirac delta scattering is implemented symbolically in the supplemental Mathematica notebook \cite{SM}.  The form of the solution is piecewise, with $\psi(x_1,x_2,t) =  \psi_{\mathrm{inc}}(x_1,x_2,t) + \psi_{\mathrm{refl}}(x_1,x_2,t)$  for $x_2 - x_1 \leq 0$ and $\psi(x_1,x_2,t) =\psi_{\mathrm{trans}}(x_1,x_2,t)$ for $x_2 - x_1 \geq 0$.
The solution follows straightfowardly from the eigenbasis

\begin{equation}
\psi_{qk}(x_1,x_2) = \begin{cases} e^{iqX+iky}+\frac{a\mu}{ik-a\mu}e^{iqX-iky} & \text{ for } y \leq 0, \\ \frac{ik}{ik-a \mu}e^{iqX+iky} & \text{ for } y\geq 0,\end{cases}
\end{equation}
where $y=x_2-x_1$ and $X=\frac{m_1x_1+m_2x_2}{m_1+m_2}$.  Here we take
\begin{align}
\psi(x_1,x_2,t) &= \int {\mathrm d}q{\mathrm d}k~ \tilde{\psi}_0(q,k) \psi_{qk}(x_1,x_2) \nonumber \\
&\times e^{-i tq^2/2(m_1+m_2)-itk^2(m_1+m_2)/2m_1m_2}, 
\end{align}
where $\tilde{\psi}_0(q,k)$ is the Fourier transform of the Gaussian initial condition
\begin{align}
\psi_0(x_1,x_2)&=\left(\left(2 \pi\right)^2  \sigma_1^2\sigma_2^2\right)^{-1/4}\exp \Big[i k_0(x_2-x_1) \nonumber\\
&\quad-{\left(\frac{m_2 y_0}{m_1+m_2}-X_0+x_1\right)^2}/{4 \sigma_1^2} \nonumber\\
&\quad-{\left(-\frac{m_1 y_0}{m_1+m_2}-X_0+x_2\right)^2}/{4 \sigma_2^2}\Big]
\end{align}

\renewcommand{\frac}[2]{\left(#1\right)/\left(#2\right)}
\renewcommand{\sqrt}[1]{\left(#1\right)^{1/2}}
The explicit form for the incident wave packet is
\begin{widetext}
\begin{dgroup*}
\begin{dmath}
\psi_{\mathrm{inc}}(X,y_,t) = \sqrt{-\frac{2m_1 m_2 \sigma _1 \sigma _2}{\pi\left(t-2 i m_1 \sigma _1^2\right) \left(t-2 i m_2 \sigma _2^2\right)}}\exp\left(\frac{\left(i t \left(X^2-2 \left(X_0-2 i k_0 \sigma _1^2\right) X+X_0^2-4 i k_0 X_0 \sigma _1^2  \\
+4 k_0 \sigma _1^2 \left(k_0 \sigma _2^2-i y_0\right)\right)+2 m_2 \left(\left(\sigma _1^2+\sigma _2^2\right) X^2-2 \left(-y \sigma _1^2+y_0 \sigma _1^2+X_0 \left(\sigma _1^2+\sigma _2^2\right)\right) X+y^2 \sigma _1^2+X_0^2 \sigma _1^2\\
+y_0^2 \sigma _1^2-2 y X_0 \sigma _1^2+2 \left(X_0-y\right) y_0 \sigma _1^2+X_0^2 \sigma _2^2-4 i y k_0 \sigma _1^2 \sigma _2^2\right)\right) m_1^3+\left(-2 k_0 \left(k_0 \left(\sigma _1^2+\sigma _2^2\right)-i y_0\right) t^2\\
+m_2 \left(3 i X^2-6 i X_0 X+4 k_0 \left(\sigma _2^2-2 \sigma _1^2\right) X+i y^2+3 i X_0^2+i y_0^2+4 y k_0 \sigma _1^2+8 k_0 X_0 \sigma _1^2+4 k_0 y_0 \sigma _1^2\\
+12 i k_0^2 \sigma _1^2 \sigma _2^2+4 y k_0 \sigma _2^2-4 k_0 X_0 \sigma _2^2-2 i y y_0\right) t+4 m_2^2 \left(\left(\sigma _1^2+\sigma _2^2\right) X^2-\left(y_0 \left(\sigma _1^2-\sigma _2^2\right)+y \left(\sigma _2^2-\sigma _1^2\right)\\
+2 X_0 \left(\sigma _1^2+\sigma _2^2\right)\right) X+X_0^2 \sigma _1^2-y X_0 \sigma _1^2+X_0 y_0 \sigma _1^2+X_0^2 \sigma _2^2-4 i y k_0 \sigma _1^2 \sigma _2^2+y X_0 \sigma _2^2-X_0 y_0 \sigma _2^2\right)\right) m_1^2\\
+m_2 \left(-4 k_0 \left(k_0 \left(\sigma _1^2+\sigma _2^2\right)-i y_0\right) t^2+m_2 \left(3 i X^2+\left(-4 k_0 \sigma _1^2+8 k_0 \sigma _2^2-6 i X_0\right) X+i y^2+3 i X_0^2+i y_0^2\\
+4 y k_0 \sigma _1^2+4 k_0 X_0 \sigma _1^2+12 i k_0^2 \sigma _1^2 \sigma _2^2+4 y k_0 \sigma _2^2-8 k_0 X_0 \sigma _2^2+4 k_0 y_0 \sigma _2^2-2 i y y_0\right) t+2 m_2^2 \left(\left(\sigma _1^2+\sigma _2^2\right) X^2\\
-2 \left(\left(y-y_0\right) \sigma _2^2+X_0 \left(\sigma _1^2+\sigma _2^2\right)\right) X+2 X_0 \left(y-y_0\right) \sigma _2^2+\left(y^2-4 i k_0 \sigma _1^2 y-2 y_0 y+y_0^2\right) \sigma _2^2\\
+X_0^2 \left(\sigma _1^2+\sigma _2^2\right)\right)\right) m_1+t m_2^2 \left(i m_2 \left(X^2-2 \left(2 i k_0 \sigma _2^2+X_0\right) X+X_0^2+4 i k_0 X_0 \sigma _2^2+4 k_0 \left(k_0 \sigma _1^2-i y_0\right) \sigma _2^2\right)\\
-2 t k_0 \left(k_0 \left(\sigma _1^2+\sigma _2^2\right)-i y_0\right)\right)}{2 \left(m_1+m_2\right){}^2 \left(t-2 i m_1 \sigma _1^2\right) \left(t-2 i m_2 \sigma _2^2\right)}\right). \label{inc}
\end{dmath}
\end{dgroup*}
\vspace{-1em}The explicit form for the reflected wave packet is
\begin{dgroup*}
\begin{dmath}
\psi_{\mathrm{refl}}(X,y_,t) = a m_1 m_2 \sqrt{\frac{\sigma _1 \sigma _2}{2 m_1^2 \sigma _1^2+i t m_1+m_2 \left(2 m_2 \sigma _2^2+i t\right)}} \exp\left(\frac{m_1 m_2 \left(m_1+m_2\right) \left(2 m_1 \sigma _1^2+i t\right) \left(2 m_2 \sigma _2^2+i t\right) a^2\\
-2 m_1 m_2 \left(2 m_1^2 \left(2 i k_0 \sigma _2^2-X+y+X_0+y_0\right) \sigma _1^2+2 m_2^2 \left(2 i k_0 \sigma _1^2+X+y-X_0+y_0\right) \sigma _2^2+2 m_1 m_2 \left(4 i k_0 \sigma _1^2 \sigma _2^2\\
+X_0 \left(\sigma _1^2-\sigma _2^2\right)+X \left(\sigma _2^2-\sigma _1^2\right)\right)+i t m_1 \left(y+y_0+2 i k_0 \left(\sigma _1^2+\sigma _2^2\right)\right)+i t m_2 \left(y+y_0+2 i k_0 \left(\sigma _1^2+\sigma _2^2\right)\right)\right) a\\
-\left(m_1+m_2\right) \left(\left(X^2-2 \left(X_0-2 i k_0 \sigma _1^2\right) X+X_0^2-4 i k_0 X_0 \sigma _1^2+4 k_0 \sigma _1^2 \left(k_0 \sigma _2^2-i y_0\right)\right) m_1^2\\
+2 \left(m_2 \left(X^2-2 X_0 X+2 i k_0 \left(\sigma _1^2-\sigma _2^2\right) X+X_0^2+4 k_0^2 \sigma _1^2 \sigma _2^2-2 i k_0 X_0 \left(\sigma _1^2-\sigma _2^2\right)\right)\\
+t k_0 \left(y_0+i k_0 \left(\sigma _1^2+\sigma _2^2\right)\right)\right) m_1+m_2 \left(2 t k_0 \left(y_0+i k_0 \left(\sigma _1^2+\sigma _2^2\right)\right)+m_2 \left(X^2-2 \left(2 i k_0 \sigma _2^2+X_0\right) X+X_0^2\\
+4 i k_0 X_0 \sigma _2^2+4 k_0 \left(k_0 \sigma _1^2-i y_0\right) \sigma _2^2\right)\right)\right)}{2 \left(m_1+m_2\right) \left(2 m_1^2 \sigma _1^2+i t m_1+m_2 \left(2 m_2 \sigma _2^2+i t\right)\right)}\right) \\
\text{$\times$erfc}\left(\sqrt{\frac{m_1 m_2 \left(2 m_1^2 \left(-2 i k_0 \sigma _2^2+X-y-X_0-y_0\right) \sigma _1^2-2 m_2^2 \left(2 i k_0 \sigma _1^2+X+y-X_0+y_0\right) \sigma _2^2\\
+a \left(m_1+m_2\right) \left(2 m_1 \sigma _1^2+i t\right) \left(2 m_2 \sigma _2^2+i t\right)+2 m_1 m_2 \left(-4 i k_0 \sigma _1^2 \sigma _2^2+X \left(\sigma _1^2-\sigma _2^2\right)+X_0 \left(\sigma _2^2-\sigma _1^2\right)\right)\\
-i t m_1 \left(y+y_0+2 i k_0 \left(\sigma _1^2+\sigma _2^2\right)\right)-i t m_2 \left(y+y_0+2 i k_0 \left(\sigma _1^2+\sigma _2^2\right)\right)\right){}^2}{2\left(m_1+m_2\right){}^2 \left(2 m_1 \sigma _1^2+i t\right) \\
\times\left(2 m_2 \sigma _2^2+i t\right) \left(2 m_1^2 \sigma _1^2+i t m_1+m_2 \left(2 m_2 \sigma _2^2+i t\right)\right)}}\right).  \label{refl}
\end{dmath}
\end{dgroup*}
The explicit form for the transmitted wave packet is
\begin{dgroup*}
\begin{dmath}
\psi_{\mathrm{trans}}(X,y_,t)= \sqrt{-\frac{2m_1 m_2 \sigma _1 \sigma _2}{\pi \left(t-2 i m_1 \sigma _1^2\right) \left(t-2 i m_2 \sigma _2^2\right)}}
\exp\left(-\frac{\left(m_1+m_2\right) \left(2 \left(m_1+m_2\right) \sigma _1^2 \sigma _2^2\\
+i t \left(\sigma _1^2+\sigma _2^2\right)\right) k_0^2}{2 m_1^2 \sigma _1^2+i t m_1+m_2 \left(2 m_2 \sigma _2^2+i t\right)}\\
+\frac{i \left(2 m_1^2 \left(-X+X_0+y_0\right) \sigma _1^2+2 m_2^2 \left(X-X_0+y_0\right) \sigma _2^2+i t m_1 y_0+i t m_2 y_0\\
-2 m_1 m_2 \left(X-X_0\right) \left(\sigma _1^2-\sigma _2^2\right)\right) k_0}{2 m_1^2 \sigma _1^2+i t m_1+m_2 \left(2 m_2 \sigma _2^2+i t\right)}-\left(\left(m_1+m_2\right){}^2 \left(X-X_0\right){}^2\right) / \left(4 m_1^2 \sigma _1^2\\
+4 m_2^2 \sigma _2^2+2 i t m_1+2 i t m_2\right)-\frac{m_1 m_2 \left(2 m_1^2 \left(2 i k_0 \sigma _2^2-X-y+X_0+y_0\right) \sigma _1^2\\
+2 m_2^2 \left(2 i k_0 \sigma _1^2+X-y-X_0+y_0\right) \sigma _2^2+2 m_1 m_2 \left(4 i k_0 \sigma _1^2 \sigma _2^2+X_0 \left(\sigma _1^2-\sigma _2^2\right)+X \left(\sigma _2^2-\sigma _1^2\right)\right)\\
+i t m_1 \left(-y+y_0+2 i k_0 \left(\sigma _1^2+\sigma _2^2\right)\right)+i t m_2 \left(-y+y_0+2 i k_0 \left(\sigma _1^2+\sigma _2^2\right)\right)\right){}^2}{2 \left(m_1+m_2\right){}^2 \left(2 m_1 \sigma _1^2+i t\right) \\
\times \left(2 m_2 \sigma _2^2+i t\right) \left(2 m_1^2 \sigma _1^2+i t m_1+m_2 \left(2 m_2 \sigma _2^2+i t\right)\right)}\right)  \left(a \exp\left(-\frac{m_1 m_2 \left(2 m_1^2 \\
\times \left(-2 i k_0 \sigma _2^2+X+y-X_0-y_0\right) \sigma _1^2-2 m_2^2 \left(2 i k_0 \sigma _1^2+X-y-X_0+y_0\right) \sigma _2^2+a \left(m_1+m_2\right) \left(2 m_1 \sigma _1^2+i t\right) \\
\times \left(2 m_2 \sigma _2^2+i t\right)+2 m_1 m_2 \left(-4 i k_0 \sigma _1^2 \sigma _2^2+X \left(\sigma _1^2-\sigma _2^2\right)+X_0 \left(\sigma _2^2-\sigma _1^2\right)\right)+t m_1 \left(i y-i y_0+2 k_0 \left(\sigma _1^2+\sigma _2^2\right)\right)\\
+t m_2 \left(i y-i y_0+2 k_0 \left(\sigma _1^2+\sigma _2^2\right)\right)\right){}^2}{2 \left(m_1+m_2\right){}^2 \left(2 m_1 \sigma _1^2+i t\right) \left(2 m_2 \sigma _2^2+i t\right) \\
\times \left(2 m_1^2 \sigma _1^2+i t m_1+m_2 \left(2 m_2 \sigma _2^2+i t\right)\right)}\right) \sqrt{\pi } \text{erfc}\left(\sqrt{-\frac{m_1 m_2 \left(2 m_1^2 \left(-2 i k_0 \sigma _2^2+X+y-X_0-y_0\right) \sigma _1^2\\
-2 m_2^2 \left(2 i k_0 \sigma _1^2+X-y-X_0+y_0\right) \sigma _2^2+a \left(m_1+m_2\right) \left(2 m_1 \sigma _1^2+i t\right) \left(2 m_2 \sigma _2^2+i t\right)\\
+2 m_1 m_2 \left(-4 i k_0 \sigma _1^2 \sigma _2^2+X \left(\sigma _1^2-\sigma _2^2\right)+X_0 \left(\sigma _2^2-\sigma _1^2\right)\right)+t m_1 \left(i y-i y_0+2 k_0 \left(\sigma _1^2+\sigma _2^2\right)\right)\\
+t m_2 \left(i y-i y_0+2 k_0 \left(\sigma _1^2+\sigma _2^2\right)\right)\right){}^2}{2\left(m_1+m_2\right){}^2 \left(2 m_1 \sigma _1^2+i t\right) \left(2 m_2 \sigma _2^2+i t\right) \\
\times \left(2 m_1^2 \sigma _1^2+i t m_1+m_2 \left(2 m_2 \sigma _2^2+i t\right)\right)}}\right) \sqrt{-\left(m_1 m_2 \left(2 m_1 \sigma _1^2+i t\right) \left(2 m_2 \sigma _2^2+i t\right)\right)/ \\
\times\left(4 m_1^2 \sigma _1^2+4 m_2^2 \sigma _2^2+2 i t m_1+2 i t m_2\right)}-1/\left(\sqrt{2} \left(2 m_1^2 \left(2 i k_0 \sigma _2^2-X-y+X_0+y_0\right) \sigma _1^2\\
+2 m_2^2 \left(2 i k_0 \sigma _1^2+X-y-X_0+y_0\right) \sigma _2^2+a \left(m_1+m_2\right) \left(t-2 i m_1 \sigma _1^2\right) \left(t-2 i m_2 \sigma _2^2\right)\\
+2 m_1 m_2 \left(4 i k_0 \sigma _1^2 \sigma _2^2+X_0 \left(\sigma _1^2-\sigma _2^2\right)+X \left(\sigma _2^2-\sigma _1^2\right)\right)+i t m_1 \left(-y+y_0+2 i k_0 \left(\sigma _1^2+\sigma _2^2\right)\right)\\
+i t m_2 \left(-y+y_0+2 i k_0 \left(\sigma _1^2+\sigma _2^2\right)\right)\right)\right)\right).  \label{trans}
\end{dmath}
\end{dgroup*}
Since these equations were generated automatically from Mathematica output, they may not be in their simplest possible forms. However, symbolic manipulation of these equations with Mathematica makes them amenable to systematic computational analysis.
\end{widetext}
}

\end{document}